\documentclass[fleqn,12pt,twoside]{article}
\usepackage{espcrc1}

\usepackage{graphicx}
\usepackage[figuresright]{rotating}

\newcommand{\AmS}{{\protect\the\textfont2
  A\kern-.1667em\lower.5ex\hbox{M}\kern-.125emS}}

\title{Dynamical study of the pentaquark antidecuplet}

\author{Fl. Stancu\address[MCSD]{Universit\'{e} de Li\`ege, 
      Institut de Physique B.5,\\
        Sart Tilman, B-4000 Li\`ege 1, Belgium} 
        \thanks{e-mail : fstancu@ulg.ac.be}}

\begin{document}

\maketitle

\begin{abstract}
Dynamical calculations are performed for all isomultiplets of
the flavour antidecuplet to which the newly discovered pentaquark
$\Theta^+$ belongs. The framework is a constituent quark model
where the short-range interaction has a
flavour-spin structure. In this model 
the lowest pentaquarks have  positive parity. 
Each antidecuplet member is
described by a variational solution with 
the Pauli principle properly taken into account.
By fitting the mass of  $\Theta^+$ of minimal content
$uudd\overline{s}$,
the mass of $\Xi^{--}$, of minimal content $ddss \overline{u}$,
is predicted at approximately 1960 MeV. The influence of the octet-antidecuplet
mixing on the masses of the $Y = 1$ and 0 pentaquarks is considered
within the same model and the role of the kinetic energy plus the hyperfine 
interaction in this
mixing is pointed out.
\end{abstract}

\vspace{1cm}

\section{Introduction}
The existence of exotic baryons containing four quarks and an antiquark in
their lowest Fock component has now a solid experimental support. 
The observation of a narrow peak at 1.54 $\pm$ 0.01 GeV/$c^2$,
called $\Theta^+$, as an $S = 1$ 
baryon resonance in the photo-production from neutron 
$\gamma n \rightarrow K^+ K^- n$ \cite{NAKANO}, has been confirmed
by several groups in various photo-nuclear reactions \cite{EXP}. 
This has been followed by the observation in $pp$ collisions \cite{NA49}
of other narrow resonances $\Xi^{--}$ and $\Xi^{0}$ at about 1862 MeV,
from which $\Xi^{--}$ is interpreted as another pure exotic member
of an SU(3) flavour antidecuplet. 
The work of Diakonov, Petrov and Polyakov \cite{DPP}
has played a particularly important role
in these discoveries. In the context of a chiral soliton model
they predicted a narrow pentaquark,
with a width of less than 15 MeV, located at the about experimentally
observed mass of $\Theta$.

At the end of the '70's,
following the observation of several signals, 
light pentaquarks were studied theoretically \cite{HOGAASEN,STROTTMAN}, but 
these signals were not confirmed. Charmed pentaquarks with strangeness,
$uuds \bar c$ and $udds \bar c$
were also predicted \cite{GI87,LI87}, 
but experimental searches carried at Fermilab
have remained inconclusive \cite{Ashery}.
These pentaquarks were introduced in the context of the
one-gluon exchange model (colour-spin interaction) and the heavy ones 
carried negative parity. On the other hand  
positive parity pentaquarks containing heavy 
flavours were proposed in the context of a pseudoscalar
exchange model (flavour-spin interaction) \cite{GR96} about ten years 
later  \cite{FS}. In this model, the lowest ones, 
$uudd \bar c$ and $uudd \bar b$, do not carry strangeness. 
Recently the H1 Collaboration at DESY \cite{H1} reported a narrow 
resonance of mass 3099 MeV, interpreted as a $uudd \bar c$ pentaquark.

The spins and parities of $\Theta^+$ and $\Xi^{--}$ are not yet known
experimentally. In this new wave of pentaquark research,
most theoretical papers take the spin equal to 1/2. The parity is 
more controversial. In chiral soliton or Skyrme models the parity
is positive \cite{DPP}. In constituent quark models it is usually positive.  
In the present approach, the parity of the
pentaquark is given by $P \ =\ {\left({-}\right)}^{\ell \ +\ 1}$,
where $\ell$ is the angular momentum associated with the relative
coordinates of the $q^4$ subsystem.  
We analyze the case where the subsystem of four light quarks 
is in a state of
orbital symmetry ${\left[{31}\right]}_{O}$
and carries an angular momentum $\ell$ = 1. Although the kinetic energy of such
a state is higher than that of the totally symmetric ${\left[{4}\right]}_{O}$
state, the ${\left[{31}\right]}_{O}$ symmetry is the most favourable
both for the flavour-spin interaction \cite{FS} and the colour-spin
interaction \cite{JM}. In the first case the statement is
confirmed by the comparison between the realistic calculations
for positive parity \cite{FS} and  negative
parity \cite{GE98}, based on the same quark model \cite{GPP}.
In Ref. \cite{FS} the antiquark was heavy, $c$ or $b$, and accordingly the 
interaction between light 
quarks and the heavy antiquark was neglected, consistent with
the heavy quark limit. In Ref. \cite{SR} an attractive spin-spin 
interaction between ${\overline s}$ and the light quarks was incorporated
and shown that a stable or narrow positive parity $uudd {\overline s}$
pentaquark can be accommodated within such a model. 
This interaction has a form that corresponds to $\eta$
meson exchange \cite{LR} and its role is to lower the 
energy of the whole system.

The purpose of this letter is to perform dynamical calculations of all
the members of the antidecuplet to which $\Theta^{+}$ and $\Xi^{- -}$ are
supposed to belong. To our knowledge this is the first attempt in this
direction. 
The present study is a natural extension of Ref. \cite{FS} where the
heavy antiquark $c$ or $b$ is now replaced by a light quark $u$, $d$ or $s$.
To describe the short range interaction
we rely on the same model \cite{GPP} as that used in \cite{FS}. That 
means that the quark-quark interaction has a flavour-spin structure
\cite{GR96}  and that the parameters are fitted to the light non-strange and
strange baryon  spectra. Moreover we assume that the quark-antiquark 
interaction is proportional to a spin-dependent operator,
but it is flavour independent, as in Ref. \cite{SR}. Its role is to introduce
the same flavour independent shift for each member of the pentaquark
antidecuplet of equal spin.
We shall fix this shift by adjusting the
mass of $\Theta^+$  to the experimental value. There is no other
free parameter in the Hamiltonian model used in this study.  For the pure
exotic $\Xi^{--}$, we predict a mass of 
1960 MeV.
For the antidecuplet members with  $Y$ = 1 and 0 we investigate 
the role of the octet-antidecuplet mixing. To some extent this study will be
a comparative one.

We search for a variational solution
of a five-body
Hamiltonian, containing a kinetic energy term, a confinement term 
and a short range (hyperfine) interaction having a flavour-spin structure.    
The SU$_F$(3) breaking is taken into account by the strange quark mass
which appears in the  mass term, in the kinetic part and in the hyperfine 
part. The latter also breaks SU$_F$(3) through the masses of the
pseudoscalar mesons exchanged among quarks.

\section{The Hamiltonian}

The Hamiltonian has the form \cite{GPP} 
\begin{equation}
H= \sum_i m_i
 + \sum_i \frac{\vec{p}_{i}^{2}}{2m_i} 
 - \frac {(\sum_i \vec{p}_{i})^2}{2\sum_i m_i} 
+ \sum_{i<j} V_{c}(r_{ij}) 
+ \sum_{i<j} V_\chi(r_{ij}) \, ,
\label{ham}
\end{equation}
with the linear confining interaction 
\begin{equation}
 V_{c}(r_{ij}) = -\frac{3}{8}~\lambda_{i}^{c}\cdot\lambda_{j}^{c} \,~ C
\, ~ r_{ij} \, ,
\label{conf}
\end{equation}
and the flavour--spin  interaction 
\begin{eqnarray}
V_\chi(r_{ij})
&=&
\left\{\sum_{F=1}^3 V_{\pi}(r_{ij})~ \lambda_i^F \lambda_j^F \right.
\nonumber \\
&+& \left. \sum_{F=4}^7 V_{K}(r_{ij})~ \lambda_i^F \lambda_j^F
+V_{\eta}(r_{ij})~ \lambda_i^8 \lambda_j^8
+V_{\eta^{\prime}}(r_{ij})~ \lambda_i^0 \lambda_j^0\right\}~
\vec\sigma_i\cdot\vec\sigma_j~.
\label{VCHI}
\end{eqnarray}
The analytic form of  $V_\gamma (r)$  
($\gamma = \pi, K, \eta$ or $\eta'$)  is 
   
\begin{equation}\label{RADIAL}
V_\gamma (r)=
\frac{g_\gamma^2}{4\pi}\frac{1}{12m_i m_j}
\{\theta(r-r_0)\mu_\gamma^2\frac{e^{-\mu_\gamma r}}{ r}- \frac {4}{\sqrt {\pi}}
\alpha^3 \exp(-\alpha^2(r-r_0)^2)\}~,
\end{equation}
with the parameters:
\begin{eqnarray}\label{PARAM}
&\frac{g_{\pi q}^2}{4\pi} = \frac{g_{\eta q}^2}{4\pi} =
\frac{g_{Kq}^2}{4\pi}= 0.67,\,\,
\frac{g_{\eta ' q}^2}{4\pi} = 1.206, &\nonumber\\ 
&r_0 = 0.43\,~ \mathrm{fm}, ~\alpha = 2.91 \,~ \mathrm{fm}^{-1}, 
C= 0.474 \, { fm}^{-2}, \, 
  m_{u,d} = 340 \,~ \mathrm{MeV}, \, m_s = 440 \,~ \mathrm{MeV}, \\
&\mu_{\pi} = 139 \,~ \mathrm{MeV},~ \mu_{\eta} = 547 \,~ \mathrm{MeV}.~
\mu_{\eta'} = 958 \, \mathrm{MeV},~ \mu_{K} = 495 \,~ \mathrm{MeV}.&
\nonumber
\end{eqnarray}
\noindent which lead to a good description of low-energy
non-strange and strange baryon spectra. Fixing the
nucleon mass at $m_N$ = 939 MeV, this parametrization gives
for example  $m_{\Delta}$ = 1232 MeV and  
$N(1440)$ = 1493 MeV.  The lowest negative 
parity states appear  at $N(1535)-N(1520)$ = 1539 MeV, i. e.
above the Roper resonance, in agreement with the experiment.


\section{The wave function Ansatz}
We start
with the $q^4$ subsystem and treat the quarks as identical particles 
in all cases. Then following Ref. \cite{FS} the orbital (O) part of 
the lowest totally antisymmetric state must carry the symmetry 
${\left[{31}\right]}_{O}$. 
In the flavour-spin (FS) coupling
scheme this state has the form
\begin{equation}  \label{STATE1}
\left.{\left|{1}\right.}\right\rangle\ =\
\left|[{31}]_O
 {\left[{211}\right]}_{C} \left[{1111}\right]_{OC}\ ;
[{22}]_{F} [22]_{S} [{4}]_{FS} 
\right\rangle
\end{equation}
which means that the wave function is totally symmetric in the flavour-spin 
space and totally antisymmetric in the orbital-colour (OC) space and that 
the $q^4$ subsystem  carries non-zero angular momentum and has zero spin. 
Then the $q^4 {\overline q}$ state is obtained by 
coupling the antiquark to the state $|1\rangle$ of Eq.(\ref{STATE1}) 
which leads to either
${\overline {10}}_F$ or to 8$_F$
and to a total spin 1/2.
To derive the orbital part 
we denote the quarks by 1, 2, 3 and 4 and the antiquark by 5 and 
introduce the internal Jacobi coordinates 

\begin{eqnarray}\label{JACOBI}
\begin{array}{c}\vec{x}\ =\ {\vec{r}}_{1}\ -\ {\vec{r}}_{2}\ , \,
\hspace{5mm} \vec{y}\ =\
{\left({{\vec{r}}_{1}\ +\ {\vec{r}}_{2}\ -\ 2{\vec{r}}_{3}}\right)/\sqrt
{3}}\\
\vec{z}\ =\ {\left({{\vec{r}}_{1}\ +\ {\vec{r}}_{2}\ +\ {\vec{r}}_{3}\ -\
3{\vec{r}}_{4}}\right)/\sqrt {6}}\ , \, \hspace{5mm} \vec{t}\ =\
{\left({{\vec{r}}_{1}\
+\ {\vec{r}}_{2}\ +\ {\vec{r}}_{3}+\ {\vec{r}}_{4}-\
4{\vec{r}}_{5}}\right)/\sqrt {10}}~.
\end{array}
\end{eqnarray}
The key issue is to construct a wave function with correct permutation
symmetry in terms of the above Jacobi coordinates.
Assuming an $s^3p$ structure for ${\left[{31}\right]}_{O}$,
the three independent ${\left[{31}\right]}_O$ states denoted 
by $\psi_i$ are  \cite{FS}

\begin{eqnarray}\label{psi1}
{\psi }_{1} = \renewcommand{\arraystretch}{0.5}
\begin{array}{c} $\fbox{1}\fbox{2}\fbox{3}$ \\
$\fbox{4}$\hspace{9mm} 
\end{array}
=
\left\langle{\vec{x}\left|{000}\right.}\right
\rangle\left\langle{\vec{y}\left|{000}\right.}\right\rangle\left
\langle{\vec{z}\left|{010}\right.}\right\rangle ~,
\end{eqnarray}

\begin{eqnarray}\label{psi2}
{\psi }_{2} = \renewcommand{\arraystretch}{0.5}
\begin{array}{c} $\fbox{1}\fbox{2}\fbox{4}$ \\
$\fbox{3}$\hspace{9mm} \end{array}
=
\left\langle{\vec{x}\left|{000}\right.}\right
\rangle\left\langle{\vec{y}\left|{010}\right.}\right\rangle\left
\langle{\vec{z}\left|{000}\right.}\right\rangle ~,
\end{eqnarray}

\begin{eqnarray}\label{psi3}
{\psi }_{3} = \renewcommand{\arraystretch}{0.5}
\begin{array}{c} $\fbox{1}\fbox{3}\fbox{4}$ \\
$\fbox{2}$\hspace{9mm} \end{array}
=
\left\langle{\vec{x}\left|{010}\right.}\right
\rangle\left\langle{\vec{y}\left|{000}\right.}\right\rangle\left
\langle{\vec{z}\left|{000}\right.}\right\rangle ~,
\end{eqnarray}
\noindent
where 
$ \left.{\left|{n\ \ell \ m}\right.}\right \rangle $
are shell model wave functions and we took the quantum number 
$m$ = 0 everywhere, for convenience. Thus each function carries an 
angular momentum $\ell $ = 1 in one of the relative coordinate,
which leads to a total parity $P$ = + 1 and a
total angular momentum $J$ = 1/2 or 3/2.
The degeneracy of these two states can be lifted by the introduction
of a spin-orbit coupling.  

The functions (\ref{psi1})-(\ref{psi3}) are used to construct a totally
antisymmetric orbital-colour state for the $q^4$ subsystem, in
agreement with (\ref{STATE1}). 
The coefficients of the resulting linear combination are fixed by group theory,
namely by the
Clebsch-Gordan coefficients of the permutation group S$_4$. In this case,
the absolute value of all three coefficients is equal to $1/\sqrt{3}$,
which means that each of the states  (\ref{psi1})-(\ref{psi3}) 
contribute with equal probability.

The pentaquark orbital wave functions are obtained 
by multiplying each $\psi_i$ by 
$\left\langle{\vec{t}\left|{000}\right.}\right\rangle$ which describes the
motion of the $q^4$ subsystem  relative to $\overline {q}$.
The wave function associated to each relative coordinate is chosen to
be a Gaussian.
This gives
\begin{equation}\label{PSI1}
{\psi }_{1}\ = 
\ \psi_0\ \ z\ {Y}_{10}\
\left({\hat{z}}\right)
\end{equation}
\begin{equation}\label{PSI2}
{\psi }_{2}\ =
\ \psi_0\ \  y\ {Y}_{10}\
\left({\hat{y}}\right)
\end{equation}
\begin{equation}\label{PSI3}
{\psi }_{3}\ = 
\ \psi_0\ \ x\ {Y}_{10}\
\left({\hat{x}}\right)
\end{equation}
where
\begin{equation}\label{PSI0}
\psi_0 = {[\frac{1}{48 \pi^5 \alpha \beta^3}]}^{1/2}
\exp\ \left[{-\ {\frac{1}{4 \alpha^2}}\ \left({{x}^{2}\ +\
{y}^{2}\ +\ {z}^{2}}\right)\ -\ {\frac{1}{4 \beta^2}}\ {t}^{2}}\right]~.
\end{equation} 
The two variational parameters are $\alpha$, the same for all internal
coordinates of the $q^4$ subsystem, $\vec{x}$, $\vec{y}$ or $\vec{z}$,
and  $\beta$, for  $\vec{t}$,  the relative coordinate
of $q^4$  to $\overline{q}$.

The algebraic structure of the  state (\ref{STATE1}) is identical to that
of Ref. \cite{CARLSON1}. The small overlap of the resulting 
$q^4 \bar q$  state with the kinematically
allowed final states could partly  explain the narrowness of 
$\Theta^+$. 



\section{Matrix elements}

The expectation values of the hyperfine interaction $ V_{\chi}$, Eq. (\ref{VCHI}),
in the flavour-spin space,
are presented in Table \ref{FOURQ}
for the three $q^4$ subsystems necessary to construct the antidecuplet. They 
are expressed in terms of the two-body radial form (\ref{RADIAL}) now denoted
as $V^{q_a q_b}_{\gamma}$ where $q_a q_b$ specifies the flavour content of 
the interacting pair. 
The SU(3)$_F$  is explicitly broken
by the quark masses and by the meson masses. By taking 
$V_{\eta}^{uu}$ = $V_{\eta}^{us}$ = $V_{\eta}^{ss}$
and $V_{\eta'}^{uu}$ = $V_{\eta'}^{us}$ = 0,
one recovers the simpler model described in Ref. \cite{CARLSON}
where one does not distinguish between the $uu$, $us$ or $ss$
pairs in the $\eta$-meson exchange.  Moreover, in Ref. \cite{CARLSON}
one takes as parameters the already integrated two-body matrix elements of 
some radial part of the hyperfine interaction, as in Ref. \cite{GR96}. 
Here  we specify a radial form,   
which allows the explicit introduction of radial
excitations at the quark level, whenever necessary.
Then, from Table \ref{FOURQ} one can easily reproduce
Table 3 of \cite{CARLSON} containing the coefficients $x_1, x_2$ and  $x_3 $, 
i. e.  the multiplicities, or the fraction of the two body matrix elements 
associated to 
$\pi, K$ and $\eta$ exchange  respectively, which appear in the expression for
the mass. The first and last row of $x_i$,
corresponding to $\Theta$ and $\Xi^{- -}$ are straightforward, inasmuch as 
their contents are $uudd \overline s$ and $ddss \overline u$ respectively.
To get the $x_i$ associated with $N_5$
and $\Sigma_5$, which we call here $N_{\overline {10}}$ and 
$\Sigma_{\overline {10}}$
respectively, one must construct the linear combinations
\begin{eqnarray}\label{COMB}
V_{\chi}(N_{\overline {10}}) = 
\frac{1}{3} V_{\chi}(uudd) + \frac{2}{3} V_{\chi}(uuds), \nonumber \\
V_{\chi}(\Sigma_{\overline {10}}) = 
\frac{1}{3} V_{\chi}(uuss) + \frac{2}{3} V_{\chi}(uuds),
\end{eqnarray}
in agreement with the flavour wave functions given in the Appendix
and the relation $V_{\chi}(uuss) = V_{\chi}(ddss)$.
Moreover, in Ref. \cite{CARLSON}, for each  exchanged meson,  one 
assumed that the radial two-body matrix elements are equal irrespective of
the angular momentum of the state,  $\ell$ = 0 or $\ell$ = 1, which we won't do.

\begin{table}
\parbox{18cm}{\caption[matrix]{\label{FOURQ}
The hyperfine interaction $ V_{\chi}  $, Eq. (\ref{VCHI}), 
integrated in the flavour-spin space, for four \\ quark subsystems.
The upper index indicates the flavour of every interacting $qq$ pair. \\ }} 
\begin{tabular}{c|c|c}
$q^4$ & $I,~ I_3$ &  $  V_{\chi}  $ \\
\hline
$uudd$ & 0,~ 0 & 30 $V_{\pi} - 2~ V^{uu}_{\eta} - 4~ V^{uu}_{\eta'} $\\
$uuds$ & 1/2,~1/2 & 15 $V_{\pi} - V^{uu}_{\eta} - 2~ V^{uu}_{\eta'}
+ 12~ V_K + 2~ V^{us}_{\eta}  - 2~ V^{us}_{\eta'}$ \\
$ddss$ & 1, -1 & $V_{\pi} + \frac{1}{3} V^{uu}_{\eta} + \frac{2}{3}V^{uu}_{\eta'} 
+ \frac{4}{3}  V^{ss}_{\eta} + \frac{2}{3} V^{ss}_{\eta'} + 20 V_K
+ \frac{16}{3} V^{us}_{\eta} - \frac{16}{3} V^{us}_{\eta'}$
\end{tabular}
\end{table}

\begin{table}
\parbox{18cm}{\caption[matrix]{\label{FIVEQ} 
Expectation values (MeV) and total energy
$E =  \sum_{n=1}^5 m_i + \langle T \rangle +
\langle V_{c} \rangle  + \langle V_{\chi} \rangle$ obtained \\
from the Hamiltonian (\ref{ham}) for various $q^4 \overline q$ 
systems. The mass $M$ is obtained from $E$ by \\
subtraction of 510 MeV in order to fit the mass of $\Theta^+$. 
The values of the variational \\
parameters $\alpha$ and $\beta$
are indicated in the last two columns.\\ }}
\begin{tabular}{c|c|c|c|c|c|c|c|c}
$q^4 {\overline q} $ &  $\sum_{n=1}^5 m_i$ & $\langle T \rangle $ 
& $\langle V_{c} \rangle $ & $\langle V_{\chi} \rangle$ & $ E $ &
$ M $ & $\alpha(fm)$ & $\beta(fm)$ \\
\hline
$uudd \overline d$ & 1700 & 1864 & 442 & -2044 & 1962 & 1452 & 0.42 & 0.92\\
$uudd \overline s$ & 1800 & 1848 & 461 & -2059 & 2050 & 1540 & 0.42 & 1.01\\
$uuds \overline d$ & 1800 & 1535 & 461 & -1563 & 2233 & 1732 & 0.45 & 0.92\\
$uuds \overline s$ & 1900 & 1634 & 440 & -1663 & 2310 & 1800 & 0.44 & 0.87\\
$ddss \overline u$ & 1900 & 1418 & 464 & -1310 & 2472 & 1962 & 0.46 & 0.92\\
$uuss \overline s$ & 2000 & 1410 & 452 & -1310 & 2552 & 2042 & 0.46 & 0.87\\
\end{tabular}
\end{table}


\begin{table}
\parbox{18cm}{\caption[matrix]{\label{ANTIDEC} 
The antidecuplet mass spectrum in MeV. \\}}
\begin{tabular}{c|c|c|c|c}
Pentaquark & $Y,~ I,~ I_3$ & Present result & Carlson et al. 
\cite{CARLSON} & Exp + GMO formula   \\
\hline
$\Theta^+$                     & 2, 0, 0       & 1540 & 1540 & 1540 \\
$N_{\overline {10}}  $         & 1, 1/2, 1/2   & 1684 & 1665 & 1647 \\
$\Sigma_{\overline {10}}$      & 0, 1, 1       & 1829 & 1786 & 1755 \\
$\Xi^{- -}$                    & -1, 3/2, -3/2 & 1962 & 1906 & 1862  
\end{tabular}
\end{table}

\section{Results and discussion}
In Table \ref{FIVEQ} we present the variational energy $E$ of the 
model Hamiltonian (\ref{ham})
resulting from the trial wave function described by Eqs. (\ref{PSI1})-(\ref{PSI0})
 for  various $q^4 {\overline q}$
systems related to the antidecuplet or the octet. 
One can see that, except for the
confinement contribution $\langle V_{c} \rangle$, all the other
terms break SU(3)$_F$, as expected:  the mass term $\sum_{n=1}^5 m_i$
increases, the kinetic energy $\langle T \rangle$ decreases and the short range 
attraction $\langle V_{\chi} \rangle$ decreases with the quark masses.
For reasons explained in the introduction, we subtract 510 MeV from the 
total energy E in order to reproduce the experimental  $\Theta^+$ mass.

For completeness, in the last two columns of Table \ref{FIVEQ}
we also indicate the values of the variational parameters
$\alpha$ and $\beta$ appearing in the radial wave function
(\ref{PSI0}) which minimize the energy of the 
systems displayed in the first column. The parameter $\alpha$  
takes values around $\alpha_0$ = 0.44 fm.
In the same quark model this is precisely the value 
 which minimizes  the
ground state nucleon  mass \cite{GE98} when the trial wave 
function is  $\phi \propto \exp[-(x^2 + y^2)/4 \alpha^2_0]$
where $\vec{x}$ and $\vec{y}$ are the first two of the 
Jacobi coordinates (\ref{JACOBI}) defined above. The quantity $\alpha_0$ 
gives a measure of the 
quark core size of the nucleon because it is its
root-mean-square radius. The parameter  $\beta$  is related to the 
relative coordinate $\vec{t}$ between the center of mass of the $q^4$
subsystem and the antiquark. It takes values about twice larger  than  $\alpha$,
which is an indication that the four quarks cluster together, whereas
$\bar q$ remains slightly separate  in contrast to certain Ansaetze 
recently promulgated in the literature.

Table \ref{ANTIDEC} reproduces the calculated antidecuplet mass spectrum 
obtained from the mass M of  Table  \ref{FIVEQ}.
The masses of $\Theta^+$ and $\Xi^{- -}$ can be read off Table \ref{FIVEQ} 
directly. The other masses are obtained from the linear combinations 
\begin{eqnarray}\label{PENTA}
M(N_{\overline {10}}) = 
\frac{1}{3} M(uudd \bar d) + \frac{2}{3} M(uuds \bar s), \nonumber \\
M(\Sigma_{\overline {10}}) = 
\frac{2}{3} M(uuds \bar d) + \frac{1}{3} M(uuss \bar s)~.
\end{eqnarray}
In comparison with  Carlson et al. \cite{CARLSON},
where the mass of $\Theta^+$ is also adjusted to the experimental value,
we obtain somewhat higher masses for $N_{\overline {10}}$,
$\Sigma_{\overline {10}}$ and $\Xi^{- -}$, the latter being about 100 MeV
above  the experimentally found mass of 1862 MeV \cite{NA49}.
This is in contrast to the 
strongly correlated diquark model of Jaffe and Wilczek \cite{JW},
where $\Xi^{- -}$ lies about 100 MeV below the experimental value.
Note that the mass of $\Theta^+$ is also fixed in that model.
In the lowest order of SU(3)$_F$ breaking, one can parametrize the result by 
the Gell-Mann-Okubo (GMO) mass formula, $M = M_{\overline {10}} + cY$. 
In the present case one obtains $M \simeq 1829 - 145~ Y$.
The fit to the measured masses of $\Theta^+$ and $\Xi^{- -}$ gives
$M \simeq 1755 - 107~ Y$. Accordingly, the masses assigned to
$N_{\overline {10}}$ and $\Sigma_{\overline {10}}$ are 1647 MeV and 1755 MeV.
They are indicated in the
last column of Table \ref{ANTIDEC}.
Starting from this fit, Diakonov and Petrov \cite{DP} analyzed 
the masses of the non-exotic members of the antidecuplet as a consequence
of the octet-antidecuplet mixing due to SU(3)$_F$ breaking.
\footnote{%
A similar analysis, but restricted to the ideal mixing
postulated by Jaffe and Wilczek \cite{JW}, has been made in Ref. \cite{OKL}.}
A new nucleon state at 1650-1690 MeV and a new $\Sigma$ at 1760-1810 MeV
have been proposed as mainly antidecuplet baryons with $Y$ = 1 and 
$Y$ = 0 respectively.
Shortly after, Pakvasa and Suzuki 
\cite{PASU} also considered the octet-antidecuplet mixing in a
phenomenological way starting from the Gell-Mann-Okubo mass formulae.
There, the 
resonance $N^*(1710)$ was taken as the $Y$ = 1 partner of 
$\Theta^+$, as in the original work of Ref. \cite{DPP}. That analysis
showed that the
range of values for the mixing angle required by the mass spectrum of
the $Y$ = 1 baryons is not consistent with the range 
needed to fit strong decays.
\footnote{%
A more extended representation mixing
including the 8, 10, $\overline {10}$,  
27, 35 and  $\overline {35}$ 
were considered in Ref. \cite{EKP}
in the context of the chiral soliton model.
The masses of $N_{\overline {10}}$ and $\Sigma_{\overline {10}}$
were predicted to be the same as those in the last column of Table
\ref{ANTIDEC}. 
The estimated range for the pure exotic pentaquarks turn out to be
1430 MeV $< M(\Theta^+) <$ 1660 MeV and 1790 MeV $< M(\Sigma^{--}) <$ 1970 MeV.}

However the recent modified PWA analysis \cite{ARNDT}
reconsiders the antidecuplet nature of $N^*(1710)$ used  
to determine the mass of $\Theta^+$ in Ref. \cite{DPP}. 
As a result, instead of $N^*(1710)$, it proposes two 
narrow resonances 1680 MeV and/or 1730 MeV,
as appropriate $Y$ = 1  partners of $\Theta^+$. This interpretation 
of the data clearly requires octet-antidecuplet mixing.

In the present model, which contains  SU(3)$_F$
breaking, the mixing appears naturally and it can be derived dynamically
starting from the Hamiltonian (\ref{ham}).
Recall that Table 3, column 3 gives the pure antidecuplet masses. 
The pure octet masses are easily
calculable using Table \ref{FIVEQ} and the octet wave functions
(see Appendix). We obtain
\begin{eqnarray}\label{OCTET}
M(N_{8}) = 
\frac{2}{3} M(uudd \bar d) + \frac{1}{3} M(uuds \bar s) = 
1568 ~\mathrm{MeV}, \nonumber \\
M(\Sigma_8) = 
\frac{1}{3} M(uuds \bar d) + \frac{2}{3} M(uuss \bar s) =
1936 ~\mathrm{MeV}.
\end{eqnarray}

The octet-antidecuplet off-diagonal matrix element, denoted by $V$, 
has only two non-vanishing contributions, one  
coming from the mass (first) term of (\ref{ham})
and associated with the overlap
of $\Phi(N_{\overline {10}})$ and $\Phi(N_8)$, or  
of $\Phi(\Sigma_{\overline {10}})$ and $\Phi(\Sigma_8)$,
and the other coming from the hyperfine interaction. Using the Appendix 
one can obtain the analytic form of $V$ as 
\begin{equation}\label{COUPLING}
V = \left\{ \renewcommand{\arraystretch}{2}
\begin{array}{cl}
\frac{2 \sqrt{2}}{3} (m_s - m_u) + 
\frac{\sqrt{2}}{3}~[S(uuds \bar s) -  S(uudd \bar d)]  
 = 166  ~\mathrm{MeV}~  
&\hspace{1.1cm} \mbox{for N} \\
\frac{2 \sqrt{2}}{3} (m_s - m_u) +
\frac{\sqrt{2}}{3}~[S(uuss \bar s) -  S(uuds \bar d)] 
= 155  ~\mathrm{MeV}~  
&\hspace{1.1cm} \mbox{for $\Sigma$} 
\end{array} \right. 
\end{equation}
where $S = \langle T \rangle + \langle V_{\chi} \rangle $.
The numerical values on the right hand side of Eq. (\ref{COUPLING})
result from the quark masses given in 
Eqs. (\ref{PARAM}) and from the
values of $\langle T \rangle $ and $\langle V_{\chi} \rangle $ 
exhibited in Table \ref{FIVEQ}.
One can see that the mass-induced breaking term is identical for
$N$ and $\Sigma$, as expected from simple  
SU(3) considerations. Its numerical value, 94.28 MeV,  
represents more than 1/2 of the total off-diagonal matrix element.

The masses of the physical states, the ``mainly octet'' $N^*$ and the 
``mainly antidecuplet'' $N_5$, result from the diagonalization of 
a 2 $\times$ 2 matrix in each case. Accordingly the nucleon solutions are
\begin{eqnarray}\label{PHYSN}
N^* =  N_{8} \cos \theta_N - N_{\overline {10}} \sin \theta_N,\nonumber \\
N_5 =  N_{8} \sin \theta_N + N_{\overline {10}} \cos \theta_N,
\end{eqnarray} 
with  the mixing angle defined by
\begin{equation}\label{ANGLE}
\tan 2 \theta_N = \frac{2 V}{M(N_{\overline {10}}) - M(N_{8})}~.
\end{equation}
The masses obtained from this mixing are 1451 MeV and 1801 MeV respectively
and the mixing angle is $\theta_N = 35.34^0$, which means that the
``mainly antidecuplet'' state $N_5$  is  67 \%
 $N_{\overline {10}}$ 
and 33 \%
$N_{8}$, and the ``mainly octet''   $N^*$ the other way round.
The latter is located in the Roper resonance mass region
1430 - 1470 MeV.  However this is a 
$q^4 \bar q$ state, i. e.  it is 
different from the $q^3$ radially excited state obtained in Ref. \cite{GPP} 
at 1493 MeV with the parameters (\ref{PARAM}) and
assigned to the Roper resonance. A mixing of the   $q^3$ and 
the $q^4 \bar q$ states could possibly be a better description  of reality. 
There is some experimental evidence that two resonances, instead of one,
separated by about 100 MeV, and located around 1440 MeV, could consistently 
describe the $\pi-N$ and $\alpha-p$ scattering in this region \cite{MORSCH},
however. 
Thus the issue of the existence of 
more than one resonance with $J^P = 1/2^+$ in the 1430- 1500 MeV 
mass range remains unsettled. 
The ``mainly antidecuplet'' solution at 1801 MeV is far from 
the higher option  of Ref. \cite{ARNDT}, at 1730 MeV, 
interpreted as the  $Y$ = 1 narrow resonance partner of $\Theta^+$.  

In a similar way we obtain two $\Sigma$ resonances, the ``mainly octet'' one
being  at 1719 MeV
and the ``mainly antidecuplet'' one at 2046 MeV.
The  octet-antidecuplet mixing angle  is
$\theta_{\Sigma} = - 35.48^0$. The lower state is somewhat above 
the experimental mass range 1630 - 1690 MeV of the 
the $\Sigma(1660)$ resonance. As the higher mass region of $\Sigma$ is
less known experimentally, it would be  difficult to make an assignement for 
the higher state.
 
The  mixing angle $\theta_N$ and  $\theta_{\Sigma}$ are nearly equal
in absolute value,  but they have
opposite signs. The reason is that  $M(N_{\overline {10}}) > M(N_8)$
while  $M(\Sigma_{\overline {10}}) < M(\Sigma_8)$. 
Interestingly, each is close to the value of the ideal mixing angle 
$\theta_N$ = 35.26$ ^0$ and $\theta_{\Sigma} = - 35.26^0$.
Only the 
relative strengths of decays and selection rules can
discriminate between mixing schemes as well as between models \cite{PASU,CD}.
This is a task for a future work.

\section{Conclusions}

In conclusion we have used a variational method, which provides upper bounds 
on the masses of all isomultiplets of the pentaquark antidecuplet.
We calculated dynamically the masses of the pure exotic pentaquarks
$\Theta^+$ and $\Xi^{--}$ and the masses of the other members of
the antidecuplet.
The model on which these calculations are based reproduces well the baryon
spectrum, when baryons are described as $q^3$ systems. 
It assumes a flavour-spin structure for the hyperfine quark-quark 
interaction   and its radial shape contains parameters
which have been fitted not only to the ground state baryons, but also
to a large number of excited states \cite{GPP}. 
In particular this interaction places the Roper resonance,
modeled as a  $q^3$ system, below the lowest negative parity baryons,
in agreement with the experiment.
However the description  of strong decays in this model is not satisfactory
(see e. g. Ref. \cite{MWP}).
Besides the $qq$ interaction a  $q \bar q$ interaction
is necessary to describe pentaquarks.  
Here we did not introduce it explicitly but relied on the 
conclusion of  Ref. \cite{SR} that 
an attractive  spin-spin interaction that operates only in
the $q \bar q$ channel can lower the $q^4 \bar q$ energy to
accommodate the $\Theta^+$.  In this way we can explain the mass shift
of -510 MeV 
necessary to reproduce the mass of $\Theta^+$.
It follows that this
flavour-independent interaction 
equally lowers all the other members of the antidecuplet and of
the octet.

But in the new light shed by the pentaquark studies, 
the usual practice of hadron spectroscopy is expected to change.
There are hints that the wave functions 
of some excited states might contain $q^4 \overline{q}$ components.
These components, if obtained quantitatively, would perhaps better explain the 
widths and mass shifts in the baryon resonances  \cite{HR}. In particular
the mass of the Roper resonance may be further shifted up or down.   
In that case the model parametrization should be revised
and more precise 
four- and five-body 
calculations should be performed.
On the other hand a full experimental confirmation of the $\Theta^+$ and
of the $\Xi^{--}$ 
resonances and more appropriate partial wave analysis 
of existing data would be
of great help in understanding the structure of pentaquarks and of
ordinary baryons.
\\



\vspace{1cm}

\centerline{\bf Appendix}

\vspace{0.5cm}
Here we give the form of one of the two independent flavour wave
functions for each isomultiplet belonging to ${\overline {10}}_F$.
It is the function where both 
pairs of quarks, 12 and 34, are in an antisymmetric state 
$\phi_{[11]}(q_a q_b) = (q_a q_b - q_b q_a)/\sqrt{2} $.
By analogy with the $q^3$ system, we shall use the label $\rho$
for all states which are antisymmetric under the permutation (12).
For $\Theta$ this wave function is straightforward
\begin{equation} 
\Phi^{\rho}(\Theta) = \phi_{[11]} (ud) \phi_{[11]} (ud) \overline s ~.
\end{equation}
The $N_{\overline {10}}$ flavour wave function is obtained from that 
of $\Theta$ by applying the $U$-spin ladder operator $U_{-}$ of SU(3).
Its normalized form becomes
\begin{equation}
\Phi^{\rho}(N_{\overline {10}}) = \frac{1}{\sqrt{3}}
\{ [\phi_{[11]} (us)  \phi_{[11]} (ud) + \phi_{[11]} (ud) \phi_{[11]} (us)] 
\overline s
+ \phi_{[11]} (ud)  \phi_{[11]} (ud) \overline d \} ~.
\end{equation}
Applying $U_{-}$ again one obtains the wave function of $\Sigma_{\overline 10}$
which is
\begin{equation} 
\Phi^{\rho}(\Sigma_{\overline {10}}) = \frac{1}{\sqrt{3}}
\{ \phi_{[11]} (us)  \phi_{[11]} (us) 
\overline s
+ [ \phi_{[11]} (us) \phi_{[11]} (ud) +  \phi_{[11]} (ud) \phi_{[11]} (us) ]
\overline d \} ~.
\end{equation}
The wave function of $\Xi^{--}$ is as simple as that of $\Theta$ but
with another quark content of course
\begin{equation}
\Phi^{\rho}(\Xi^{--}) = \phi_{[11]} (ds) \phi_{[11]} (ds) \overline u ~.
\end{equation}
In these functions the normal order of particles 1234 is understood.
In each case one can get the other
linear independent function in the flavour space, $\Phi^{\lambda}$,
with the quark pairs 12 and 34 in a symmetric state, 
$\phi_{[2]}(q_a q_b)
= (q_a q_b + q_b q_a)/\sqrt{2} $ ( $q_a \neq q_b)$
or  $\phi_{[2]}(q_a q_a) =q_a q_a$, by applying the permutation
(23) to the above corresponding function 
(see e. g. \cite{book}). 
For example we have
\begin{equation} 
\Phi^{\lambda}({\Theta}) = \sqrt{\frac{1}{3}}
[ \phi_{[2]} (uu) \phi_{[2]} (dd)
+ \phi_{[2]} (dd) \phi_{[2]} (uu)
- \phi_{[2]} (ud) \phi_{[2]} (ud) ] \overline s ~.
\end{equation}
Both the $\Phi^{\rho}$ and $\Phi^{\lambda}$ functions are necessary
in the calculation of the matrix elements of the hyperfine interaction.

In the same notation, the $N_8$ and $\Sigma_8$
the flavour octet wave functions, 
antisymmetric under the permutation (12) are
\begin{equation}
\Phi^{\rho}(N_{8}) =\frac{1}{\sqrt{6}}
 [\phi_{[11]} (us)  \phi_{[11]} (ud) + \phi_{[11]} (ud) \phi_{[11]} (us)] 
\overline s 
- \sqrt{\frac{2}{3}} \phi_{[11]} (ud)  \phi_{[11]} (ud) \overline d  ~.
\end{equation}
\begin{equation} 
\Phi^{\rho}(\Sigma_8) = \sqrt{\frac{2}{3}}
 \phi_{[11]} (us)  \phi_{[11]} (us) 
\overline s
- \frac{1}{\sqrt{6}} 
[ \phi_{[11]} (us) \phi_{[11]} (ud) +  \phi_{[11]} (ud) \phi_{[11]} (us) ]
\overline d  ~.
\end{equation}
\vspace{1cm}

\centerline{\bf Acknowledgment}
\vspace{0.5cm}
I am indebted to Dan Riska and 
for useful comments on the manuscript and to Veljko Dmitra\v sinovi\' c 
for fruitful discussions.

\vspace{1cm}



\end{document}